\documentclass[dvips]{article}

\usepackage{icrc2011}

\title{The search for galactic dark matter clump candidates with Fermi and MAGIC}

\newcommand{\etal}{\MakeLowercase{\textit{et al. }}} 
\shorttitle{Nieto \etal The search for DM clumps with Fermi and MAGIC}

\authors{Nieto D.$^{1}$, Aleksi\'{c} J.$^{2}$,
Barrio J.A.$^{1}$, Contreras J.L.$^{1}$, Doro M.$^{3}$, Lombardi
S.$^{4}$, Mirabal N.$^{1}$, Moralejo A.$^{2}$, Pardo S.$^{1}$, Rico
J.$^{2,5}$, Zandanel F.$^{6}$, on behalf of the MAGIC Collaboration}
\afiliations{$^1$Universidad Complutense de Madrid, Spain\\
  $^2$Instituto de F\'{i}sica de Altas Energ\'{i}as, Spain\\
  $^3$Universidad Aut\'{o}noma de Barcelona, Spain\\
  $^4$Universit\'{a} degli Studi di Padova, Italy\\
  $^5$Instituci\'{o} Catalana de Recerca i Estudis Avan\c{c}ats, Spain\\
  $^6$Instituto de Astrof\'{i}sica de Andaluc\'{i}a, Spain\\}
\email{nieto@gae.ucm.es}

\abstract{
We present a systematic search for potential dark matter clumps in our
Galaxy among the 630 unassociated sources included in the LAT 1-year
Point Source Catalog.  Assuming a dark matter particle that generates
observable gamma-ray photons beyond the Fermi energy range through
self-annihilation, we compile a list of reasonable targets for the
MAGIC Imaging Atmospheric Cherenkov Telescopes. In order to narrow
down the origin of these enigmatic sources, we summarize ongoing
multiwavelength studies including X-ray, radio, and optical
spectroscopy. We report on observations of two of these candidates
using the MAGIC Telescopes. We find that the synergy between Fermi and
Cherenkov telescopes, along with multiwavelength observations, could
play a key role in indirect searches for dark matter.
}

\begin{document}
\maketitle

\section{Introduction}

The concordance cosmological model, thoroughly validated by
measurements, requires 83\% of the total mass density in the Universe
to be non-baryonic \cite{Komatsu:2011a}. Thus, the identification of
this so-called \emph{dark matter} (DM) is one of the most relevant
issues in Physics today. Assuming that DM is composed by weakly
interacting massive particles (WIMP), which could annihilate or decay
into standard model particles, its nature can be unraveled by the
detection of these by-products. This is the principle of indirect
detection searches carried out in the $\gamma$-ray regime.

A $\gamma$-ray signal from DM annihilation would be characterized by a
very distinctive spectral shape due to features such as annihilation
lines \cite{Bertone:2009a} and internal bremsstrahlung
\cite{Bringmann:2008a}, as well as a characteristic cut-off at the DM
particle mass. In order to shed light over the nature of the DM
constituent the detection of several sources sharing the same DM-like
spectrum is mandatory, since the DM spectrum must be universal
\cite{Pieri:2009a}. Astrophysical regions where high DM density is
foreseen are the best candidates to search for DM originated
$\gamma$-ray emission. No DM signal has been detected so far in any of
the most promising targets, including dwarf spheroidal galaxies
\cite{Aleksic:2011a}, galaxy clusters \cite{Aleksic:2010a} or the
Galactic Center \cite{Abramowski:2011b}.

Yet, there exist other possible regions of high DM density. Most
recent cosmological N-body high-resolution simulations
\cite{Springel:2008b} indicate that DM halos should not
be smooth but must exhibit a wealth of substructure on all resolved
mass scales \cite{Diemand:2008a}. These subhalos could be
too small to have attracted enough baryonic matter to start
star-formation and would therefore be invisible to past and present
astronomical observations. Overdensities or clumps are foreseen into
these subhalos which can be nearby in our galaxy and therefore bright
at $\gamma$-rays \cite{Pieri:2008a}. Also DM high density regions can
develop around intermediate massive black holes where a rather peaked
$\gamma$-ray emission is predicted \cite{Bertone:2009b}. These clumps
would most probably only be visible at very high energies (VHE) and
therefore may not have shown up in any catalog yet.

Since $\gamma$-ray emission from DM annihilation is expected to be
constant, DM clumps would pop-up in all-sky monitoring programs
\cite{Kamionkowski:2010a}. This can be best provided by the
Fermi satellite telescope\footnote{http://fermi.gsfc.nasa.gov/} as
\emph{unassociated Fermi objects} (UFOs) not detected at any other
wavelengths. Very likely, the distinct spectral cut-off at the DM
particle mass is located at too high an energy (see, \emph{e.g.} the
\emph{neutralino} mass lower limits in \cite{Nakamura:2010a}) to be
measurable by Fermi within reasonable time and can only be
limited by IACT observations.

\section{Candidates Search}
The First Fermi-LAT Catalog (1FGL) contains 1451 high energy
$\gamma$-ray sources detected by the LAT instrument after the first 11
months of the science phase of the mission \cite{Abdo:2010a}. For each
source, positional and spectral information are provided as well as
identification or possible associations with cataloged sources at
other wavelengths. Although Fermi-LAT has a good angular
resolution, a firm identification based on positional coincidence
alone is not always feasible. Thus, 630 sources in the 1FGL lack any
clear association. These are the so-called unassociated Fermi objects
(UFOs), a population among which DM clumps might be represented
\cite{Buckley:2010a}.

\subsection{Selection Criteria}
In order to extract possible DM clump candidates out of the 1FGL UFOs
the following selection criteria were required:
\vspace{-6mm}
\begin{itemize}
\item To lay outside the Galactic Plane.
\end{itemize}
\vspace{-3mm} 
A noteworthy fraction of galactic baryonic objects are
found in the Galactic Plane, unlikely the galactic dark matter
substructures whose galactic latitude distribution is homogeneous
\cite{Diemand:2008a}. Source association in a very
crowded environment is more difficult for the Fermi association
algorithms. Thus associations due to an excess of candidates are
more likely. Moreover, the galactic diffuse $\gamma$-ray background is
much stronger at low galactic latitudes. Consequently, UFOs with
galactic latitudes $|b|<10^{\circ}$ were discarded.
\vspace{-3mm}
\begin{itemize}
\item To be a hard source.
\end{itemize}
\vspace{-3mm} 
The expected spectrum from WIMP annihilation essentially follows the shape of the annihilation photon
yield, which has been shown to be hard up to the WIMP mass cut-off
\cite{Bertone:2006a,Cembranos:2010a}. Additionally, 1FGL sources
showing hard spectra are more likely to be detected by IACTs beyond the Fermi
upper energy threshold. Therefore only hard sources were
selected, meaning that 1FGL sources whose spectral fitting power law
index was above 2 were discarded.
\vspace{-3mm}
\begin{itemize}
\item To be non-variable.
\end{itemize}
\vspace{-3mm} The photon flux from dark matter annihilation must be
constant, thus variable sources were rejected. Sources
with a 1FGL \emph{variability index} showing the lightcurve to
deviate from a flat one were discarded.
\vspace{-3mm}
\begin{itemize}
\item To follow a power law spectra.
\end{itemize}
\vspace{-3mm} DM spectra show prominent cut-offs at the DM particle
mass. Sufficiently away from the cut-off the spectra can be well
described by a power law (see \emph{e.g.} the asymptotic behavior of
spectra in \cite{Bertone:2006a}). Thus, assuming that the cut-off lays
beyond Fermi energy range, sources departing from a power law
spectral fit (information which is provided by the 1FGL
\emph{curvature index}) were rejected.
\vspace{-3mm}
\begin{itemize}
\item Not to have possible counterparts.
\end{itemize}
\vspace{-3mm} 
An extensive and independent search for possible associations was
performed for each UFO through the NASA's High Energy Astrophysical
Archive\footnote{http://heasarc.gsfc.nasa.gov/}.  The main
astronomical catalogs and missions archives, from $\gamma$-ray to
radio, were explored around the sources 1FGL nominal positions with a
20' conservative search radius corresponding to twice Fermi PSF at 10
GeV \cite{Burnett:2009a}, and UFOs with possible counterparts were
discarded.  Also Swift-XRT data from several high galactic
latitude UFOs \cite{Donato:2010a} were made public recently. After
analyzing these data, UFOs containing Swift-XRT X-ray sources
within their Fermi error contour were consequently discarded.  The
purpose of this search was not to associate nor to identify
counterparts for 1FGL sources, but to conservatively discard objects
whose Fermi $\gamma$-ray flux could be eventually attributed to an
already detected source.\\ After all these criteria, the candidate
search finally provided 10 possible DM clumps out of the 630 initial
UFOs.

\subsection{IACTs Detection Prospects}

A signal detection in the IACT context is defined as a more than
5$\sigma$ deviation of the excess events over the background
events. If the total number of observed events is expressed in terms
of their rates as $N_{on}=(R_{exc}+R_{bkg})t$ and
$N_{off}=\kappa$$R_{bkg}t$, were $t$ is the observation time, and the
on-off ratio is assumed to be $\kappa = 1$, the detection time can
then be estimated working out Eq. 5 from Li and Ma~\cite{Li:1983a}.
The excess rate $R_{exc}$ over a certain energy threshold $E_{th}$ can
be computed from the effective area of the instrument, and the
differential spectrum of the source. The characteristic background
rate of the instrument is computed from Monte Carlo simulations.

For this work, the MAGIC
Telescopes\footnote{http://wwwmagic.mppmu.mpg.de/} effective area and
background rate were considered. The very high energy UFO spectra were
directly extrapolated from the 1FGL Catalog, evaluating also the
impact of the uncertainties in the Fermi spectral parameters over the
detection time. The adopted $E_{th}$ was 100 GeV, a conservative one
already achieved by MAGIC single telescope observations.

Fermi data were studied for all these 10 sources using the
latest version of the Fermi \texttt{ScienceTools}
\cite{FERMItools:2011}. The total number of high energy
photons (HE, $E_{\gamma}>10~$GeV) is a determinant quantity since it
provides an evidence of the validity of the Fermi spectra
extrapolation. Therefore, HE photons from a circular region of 1.5
times Fermi PSF radius (0.15$^{\circ}$) were extracted in order
to get the events likely to have been emitted by the source (the
contribution from diffuse HE $\gamma$-ray background in each source
position was estimated from actual data to be almost
negligible).

Finally, the estimated detection time, and the list of
Fermi HE photons, were used to sort the 10 candidates attending
to their feasibility of detection: sources were ordered as a function
of their estimated detection time, although the effect of the spectral
parameters uncertainties was taken into consideration, giving lower
priority to those whose uncertainties were larger; in case of similar
estimated detection times, sources with a larger Fermi HE
photons population were given higher priority.

\section{MAGIC Observations}

MAGIC consists of a system of two telescopes operating in stereoscopic
mode since fall 2009 at the Canary Island of La Palma ($28.8^{\circ}$
N, $17.8^{\circ}$ W, 2200 m a.s.l.). Only 6 out of the 10 selected DM
clump candidates can be observed from MAGIC latitude under reasonable
zenith angle conditions. So far, the two best candidates, namely, 1FGL
J2347.3+0710 and 1FGL J0338.8+1313, have been observed under dark
night conditions and lowest zenith angle range possible. Both
conditions are needed when the sensitivity at low energies is
pursued. The sources were surveyed in false tracking mode
\cite{Fomin:1994a}.  In the two cases, data were analyzed in the
\texttt{MARS} analysis framework by means of the
standard stereoscopic analysis routines
\cite{Aleksic:2011c}. Contemporaneous Crab Nebula data were used to
verify the proper performance of the telescopes and analysis routines.

\begin{figure}[t!]
\centering
\includegraphics[width=1\linewidth]{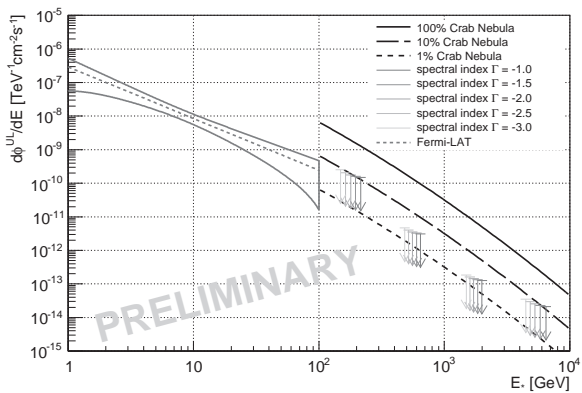}
\caption[1FGL J2347.3+0710 differential spectrum]{1FGL J2347.3+0710
  Fermi-LAT differential spectrum and preliminary MAGIC differential
  spectrum upper limits. MAGIC Crab Nebula spectrum
  \cite{Albert:2007a} is depicted as reference.}
\label{fig:diffUL_1FGLJ2347}
\end{figure}

\subsection{1FGL J2347.3+0710 Observations}

The observation of 1FGL J2347.3+0710 were performed during October and
November 2010. The zenith angle window ranged from $21.5^{\circ}$ to
$30.0^{\circ}$. The total exposure time was 13.3 h. After data quality
selection the exposure time reduced down to 8.3 h.\\ No signal was
found over the background. Considering an energy threshold of 100 GeV
the number of excess events was $N_{exc}(>100~$GeV$)=98\pm86$ events
which translates into a significance of $1.1\sigma$, computed using
Eq. 17 from Li and Ma \cite{Li:1983a}. Consequently, we derived upper
limits (ULs) to the differential and integral spectra following the
prescriptions from \cite{Aleksic:2011a}. The integral ULs for
different energy thresholds and power law spectra are found in
Table~\ref{tab:intUL}.  The differential ULs are presented in
Fig.~\ref{fig:diffUL_1FGLJ2347}, together with the corresponding 1FGL
Catalog Fermi spectrum and its error band (computed as in
\cite{Abdo:2010c}).

\subsection{1FGL J0338.8+1313 Observations}
\begin{figure}[t!]
\centering
\includegraphics[width=1\linewidth]{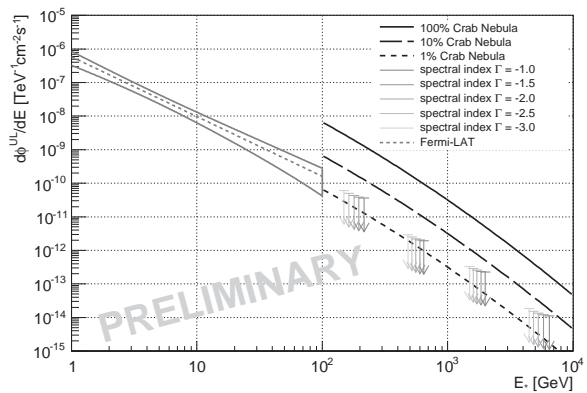}
\caption[1FGL J0338.8+1313 differential spectrum]{1FGL J0338.8+1313
  Fermi-LAT differential spectrum and preliminary MAGIC differential
  spectrum upper limits. MAGIC Crab Nebula spectrum
  \cite{Albert:2007a} is depicted as reference.}
\label{fig:diffUL_1FGLJ0338}
\end{figure}

In the case of 1FGL J0338.8+1313 the observations were performed from
December 2010 to January 2011. The zenith angle window covered the
interval from $15.5^{\circ}$ to $30.5^{\circ}$, again ensuring a low
energy threshold. Data were taken for a total observation time of 15.5
h which reduced to 10.7 h after data quality selection.  As in the
previous case, no signal was detected over the background. The excess
events above an energy threshold of 100 GeV were
$N_{exc}(>100~$GeV$)=-81\pm84$, producing a significance of
$-1.0\sigma$. The integral ULs, extracted as already described, can be
found in Table~\ref{tab:intUL}. MAGIC differential ULs as well as
Fermi spectrum with its error band are found in
Fig.~\ref{fig:diffUL_1FGLJ0338}

\begin{table*}[ht!]
\centering
\vspace{-3mm}
\caption[Integral upper limits]{Integral upper limits.}
\resizebox{0.9\linewidth}{!}{
  \begin{small}
  \begin{tabular}{ccccccccc}
    \hline
    E$_0$ & $N_{\mbox{\tiny{ON}}}/N_{\mbox{\tiny{OFF}}}$ &  $\sigma_\mathrm{Li,Ma}$ & $N^{UL}_{exc}$ & \multicolumn{5}{c}{$\Phi^{UL}$} \\ 
    $[$GeV$]$ & & & 95\% C.L. & \multicolumn{5}{c}{$\times 10^{-11} [$cm$^{-2}$s$^{-1}]$} \\
    & & & & $\Gamma=-1.0$ & $\Gamma=-1.5$ & $\Gamma=-2.0$ & $\Gamma=-2.5$ & $\Gamma=-3.0$ \\
    \hline
    1FGL J2347.3+0710 & & & & & & & & \\
    100 & 3744/3646 &  1.14 & 376 & 16.1 & 20.8 & 26.8 & 31.0 & 32.6 \\
    250 &   369/372 & -0.11 &  62 &  2.4 &  2.6 &  3.0 &  3.2 &  3.3 \\
    500 &     77/92 & -1.15 &  18 &  0.6 &  0.6 &  0.7 &  0.7 &  0.7 \\
    1000&     18/27 & -1.34 &   9 &  0.3 &  0.3 &  0.3 &  0.3 &  0.3 \\
    \hline
    1FGL J0338.8+1313 & & & & & & & & \\
    100 & 3494/3575 & -0.96 & 119 & 4.0 & 5.1 & 6.6 & 7.6 & 8.0\\
    250 &   346/350 & -0.15 &  59 & 1.8 & 2.0 & 2.2 & 2.4 & 2.4\\
    500 &     81/82 & -0.07 &  30 & 0.9 & 0.9 & 0.9 & 0.9 & 1.0\\
    1000&     18/19 & -0.16 &  14 & 0.4 & 0.4 & 0.4 & 0.4 & 0.4\\
    \hline
  \end{tabular}
  \end{small}
}
  \label{tab:intUL}
\end{table*}

\section{Discussion \& Conclusions}
A dedicated search designed to select possible DM clump candidates out
of the 1FGL Catalog has been presented, concluding with 10 candidates
out of the 630 UFOs. After studying the prospects of detection for
each of these 10 sources, the two best candidates were observed by the
MAGIC Telescopes.\\Although no very high energy $\gamma$-ray
signal was detected for any of them, competitive upper limits to the
differential and integral spectra were obtained.\\It can be seen from
Fig.~\ref{fig:diffUL_1FGLJ2347} that a direct extrapolation of 1FGL
J2347.3+0710 Fermi spectrum above 400 GeV is ruled out by
MAGIC observations, meaning that some kind of cut-off or
spectral curvature may be taking place at energies between 100 and 400
GeV. In the case of 1FGL J0338.8+1313, as illustrated in
Fig.~\ref{fig:diffUL_1FGLJ0338}, one can conservatively rule out a
direct extrapolation of Fermi spectrum above 200 GeV. This fact
suggests a possible curvature or cut-off at Fermi high energy
range.\\ Nonetheless, these conclusions should be taken \emph{cum grano
  salis} since they rely on 1FGL spectral information, and must be
consequently revisited once the second version of the Fermi
Catalog (2FGL) is released.

We expect the synergy between deeper MAGIC observations and the
incoming 2FGL Catalog will help us to reveal the actual spectral
nature of these two enigmatic objects and the next-to-come in future
searches of DM clump candidates.

\section{Acknowledgments}
We would like to thank the Instituto de Astrof\'{\i}sica de Canarias
for the excellent working conditions at the Observatorio del Roque de
los Muchachos in La Palma.  The support of the German BMBF and MPG,
the Italian INFN, the Swiss National Fund SNF, and the Spanish MICINN
is gratefully acknowledged. This work was also supported by the Marie
Curie program, by the CPAN CSD2007-00042 and MultiDark CSD2009-00064
projects of the Spanish Consolider-Ingenio 2010 programme, by grant
DO02-353 of the Bulgarian NSF, by grant 127740 of the Academy of
Finland, by the YIP of the Helmholtz Gemeinschaft, by the DFG Cluster
of Excellence ``Origin and Structure of the Universe'', and by the
Polish MNiSzW grant 745/N-HESS-MAGIC/2010/0.  N.M. gratefully
acknowledges support from the Spanish MICINN through a Ram\'on y Cajal
fellowship.

\bibliographystyle{h-physrev}
\bibliography{refs} 

\clearpage

\end{document}